# A Completely Blind Channel Estimation Technique for OFDM Using Generalized Constellation Splitting and Modified Phase-Directed Algorithm

Mr. Sameera Bharadwaja H.
Department of Electronics and Computer Engineering
Indian Institute of Technology Roorkee
Roorkee, Uttarakhand, India – 247 667
Email: sameerabharadwaja@gmail.com

Dr. D. K. Mehra
Department of Electronics and Computer Engineering
Indian Institute of Technology Roorkee
Roorkee, Uttarakhand, India – 247 667
Email: dkmecfec@iitr.ernet.in

***Abstract*- The problem of blind channel estimation for SISO-OFDM systems using second-order statistics (SOS) is addressed. A comparison of two prominent SOS-based techniques: subspace-decomposition and precoding-induced correlation-averaging techniques in terms of MSE performance is presented. The drawback of these methods is that they suffer from a complex-scalar estimation ambiguity which is resolved by using pilots or reference symbols. By using pilots the whole purpose of blind techniques is contradicted. We propose an algorithm to resolve this ambiguity in blind manner using generalized constellation-splitting and modified phase-directed algorithm. The performance of the proposed scheme is evaluated via numerical simulations in MATLAB environment.**

## I. INTRODUCTION

In a digital communication link, prior to the coherent detector block, the incoming information symbol needs to be equalized of imperfect channel variations to prevent erroneous decisions [1]. The channel impulse response (CIR) knowledge is crucial to ensure acceptable performance of equalizer. In practice, the CIR is estimated by using training. But, for time-varying channels, training based methods are unsuitable because of high bandwidth overhead introduced [2]. The blind techniques (no training) have been researched during past three decades [3, 4, 5 and references therein summarises almost all blind approaches proposed up to last decade]. Since training overhead is eliminated; higher spectral efficiency and higher information rates can be achieved.

The early blind approaches are based on utilization of higher-order statistics (HOS). Extracting information from HOS involves non-linear processing (computationally complex) and has low convergence rate. This compromise on convergence rate is unacceptable in applications wherein the channel is time-varying (mobile channels), data rates are high or the data is sent in short bursts (TDMA). In mid 1990's and onwards, the SOS-based approaches were proposed for single-carrier systems. These methods have relatively low processing demands and good convergence rates. In the early part of past decade, multi-carrier transmission, especially OFDM have gained much importance [6]. Further, OFDM has been adopted as a standard in DAB, DVB (HDTV), Wireless LAN (IEEE 802.11), WiMAX (IEEE 802.16), 3 GPP Long-Term Evolution (LTE) etc.

Over time, blind subspace-based OFDM channel estimation methods have been proposed [7, 8] which can estimate the channel up to a constant complex-scalar ambiguity. Recently, precoding technique at the transmitter has been adopted to introduce certain correlation among the subcarriers in OFDM frame [9, 10]. This correlation information is used to estimate channel up to a constant phase ambiguity. Further, by precoding, estimation of arbitrary channels using SOS is possible, thereby making precoding-based methods attractive over subspace-based methods.

The phase ambiguity fundamental to most of the SOS-based methods are resolved in [7-10] via pilot/ reference symbols. By using pilots, whole purpose of blind methods is contradicted. The thumb rule is that, more the information about the source known at the receiver, better the estimation accuracy is. According to the author's knowledge, very little work on resolving this ambiguity blindly using SOS-based methods has been done. A very few handful of references are available in the literature. As far as SISO-OFDM systems are concerned, some characteristics like channel coding [11], finite alphabet property of the source [12], utilizing asymmetric constellation [13], mixed-order modulation [14] and source statistics information are employed.

In some methods, it is assumed that the source symbols are taken from a finite alphabet set [12]. For example, consider a system in which PSK modulation is used. This scheme has constant modules property, with the symbol phase chosen from a finite known set. This source property can be used to resolve the phase ambiguity. Some of known algorithms [12] are minimum-distance (MD), phase-directed (PD) and decision-directed (DD) approaches. MD aims to minimize the Euclidian distance measure and involves searching over all possible values of channel vectors; PD is an iterative process in which ambiguity is resolved by searching over phase values and DD consists of equalizing $k^{th}$ subcarrier and projecting it on to a valid symbol from the set. This equalized symbol is then used to estimate the channel co-efficient. Since PSK modulation which has constant modulus property is used, this method is not applicable for any other mapping schemes like PAM and QAM.

Some approaches wherein, in the source constellation, one or more symbol point is skewed so as to get an asymmetrical arrangement [13] have been proposed. The phase relationship information due to this variation is utilized for phase correction. The disadvantage is that, since asymmetry is induced in the source constellation, the dc level in an OFDM data frame is non-zero in statistical sense. This may cause significant dc power dissipation and hence not feasible for power-limited systems like handheld and mobile transceiver devices.

As far as PSK is concerned, mixed-order modulation among alternate subcarriers can be employed as proposed by Necker [14]. This type of mapping induces a unique phase relationship among alternate subcarriers that can be employed in phase-correction. For instance, a 3-PSK and 4-PSK can be combined and assigned among alternate subcarriers in a given OFDM frame. Since, the symbol phases are non-overlapping, phase correction is possible. But, Mary-PSK mapping with $M \neq 2^n$ ($n \in Z^+$) means non-binary coding. For example, 3-PSK can be generated by mapping a ternary symbol set $S = \{0, 1, 2\}$ on to three phase values. This means that the quantizer design has to be altered which leads to encoding/ decoding design issues. Further, this approach uses ML-type estimator to obtain first approximation of the channel. ML estimator demands high implementation complexity and hence is not practically feasible.

The problems of dc-level, synchronization, non-uniform quantization/ coding and non-applicability to non-constant modulus mapping have been overcome by an algorithm proposed by Sameera Bharadwaja H. and D. K. Mehra [15] for SISO-OFDM systems with PAM mapping. The channel is estimated in frequency-domain up to a constant phase ambiguity factor by precoding-based technique. A constellation-splitting technique applicable only for PAM mapping is used to provide the required side-information to the phase-estimator-corrector block for phase-ambiguity removal. This paper proposes a blind algorithm to resolve the ambiguity by using received signal SOS in blind manner via generalized constellation-splitting and a modified-phase-directed algorithm which is applicable for any modulation schemes.

The rest of the paper is organized as follows: Section II presents the system model. Section III presents the comparison between the subspace- and precoding-based techniques wherein the ambiguities are resolved by using pilots. Section IV presents the completely blind algorithm. In Section V, the simulations results are provided to show the performance comparison of the proposed method with respect to its semi-blind counterparts (Section II). Conclusion is given in Section VI. Standard mathematical notations and MATLAB notations are used.

## II. SYSTEM MODEL

Fig. 1 shows a model for the baseband SISO-OFDM system. Each OFDM frame consists of $N$ subcarriers. The channel is assumed to be frequency-selective represented by vector $\boldsymbol{h} = [h_0, \ldots, h_L]^T$. Let the $n^{th}$ block of the frequency-domain information symbols be written as, $\boldsymbol{d}(n) = [d(n, 0), \ldots, d(n, N-1)]^T$. Each symbol $d(n,k)$ is taken from a baseband frequency-domain signal constellation (map). The precoder block multiplies the incoming OFDM frame by a predefined matrix $\boldsymbol{W}$ to yield $\boldsymbol{x}(n)$. The OFDM modulator, applies an N-point IFFT to this block, and inserts the CP in front of the IFFT output vector, which is a copy of the last $L$ samples of the IFFT output. This results in the time-domain sample vector of the $n^{th}$ OFDM symbol written as $\boldsymbol{s}(n) = [s(n, N-P), \ldots, s(n, N-1), s(n, 0), \ldots, s(n, N-1)]^T$.

As far as subspace-based technique is concerned, the precoding matrix is assumed to be $\boldsymbol{I}_N$. Collecting two consecutive time-domain received OFDM blocks $\boldsymbol{r}_{cp}(n)$ and $\boldsymbol{r}_{cp}(n\text{-}1)$ to form a composite block $\bar{\boldsymbol{r}}(n)$, the following time-domain input-output relation can be formed which is used for subspace decomposition [7, 16]:

$$\bar{\boldsymbol{r}}(n) = \mathbf{H}(h)\bar{\boldsymbol{s}}(n) + \bar{\boldsymbol{n}}(n) \qquad (1)$$

where, $\boldsymbol{H}(h)$ is the time-domain channel convolutional matrix of dimension $(2N+L) \times 2N$; $\bar{\boldsymbol{r}}(n)$, $\bar{\boldsymbol{s}}(n)$ and $\bar{\boldsymbol{n}}(n)$ are the composite time-domain output, input and the noise blocks of dimension $(2N+L)\times1$, $2N\times1$ and $(2N+L)\times1$ respectively. Let $\boldsymbol{H} = [H_0, \ldots, H_{N-1}]^T = DFT(\boldsymbol{h})$ be the channel frequency response vector of dimension $N\times1$.

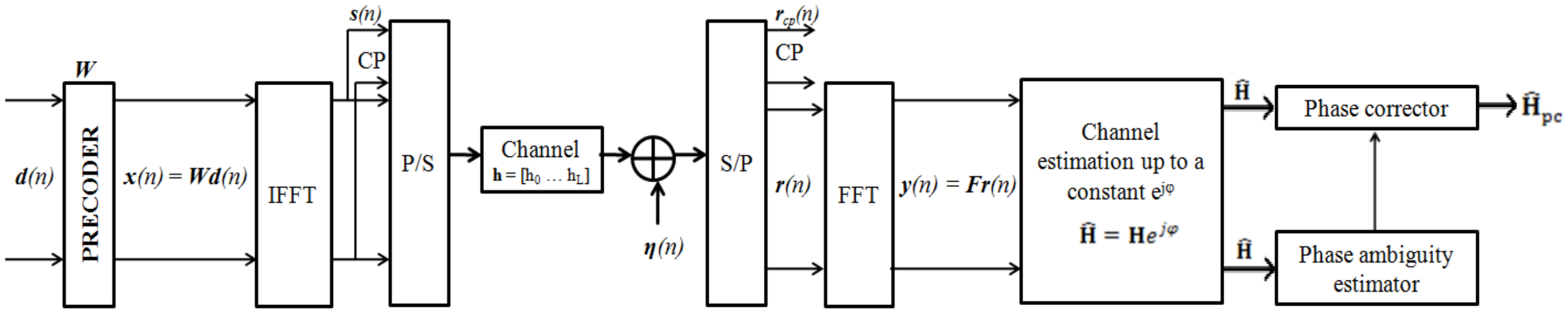

Figure 1: Discrete baseband equivalent SISO-OFDM channel model

The following frequency-domain input-output relation can be formed which can be used for precoding-based method [10]

$$\boldsymbol{y}(n) = \widetilde{\boldsymbol{H}}\boldsymbol{W}\boldsymbol{d}(n) + \widetilde{\boldsymbol{n}}(n) \tag{2}$$

where, $\boldsymbol{W}$ is linear frequency-domain non-redundant precoding matrix and $\widetilde{\boldsymbol{H}}$ is the frequency-domain diagonal channel matrix whose $(k, k)^{th}$ diagonal entry is the $k^{th}$ element of $\boldsymbol{H}$. As mentioned earlier, these techniques can, in general, estimate the channel up to a complex-scalar ambiguity. The ambiguity-estimator and corrector block can resolve this ambiguity either by using pilots or in blind manner as applicable.

## III. A COMPARISON OF SUBSPACE- AND PRECODING-BASED TECHNIQUES

The performance comparison of subspace-based [16] and precoding-induced correlation-averaging techniques [10] for SISO-OFDM, wherein the estimation ambiguities are resolved using pilots, in terms of mean square error (MSE) over different baseband modulation (mapping) schemes and channel power delay profiles are presented in this section. A MSE performance plots versus Length of OFDM blocks and SNR (dB) values are shown in Fig. 2. The parameters used for simulation purpose are given below as follows:

- The discrete baseband equivalent FIR channel impulse response is assumed to be static over the interval of estimation process with length ($L$) equals to *2*. The channel coefficients are assumed to be complex zero-mean Gaussian distributed random variable with variances given according to either of the following power-delay-profiles (PDP) [17]:
  - Exponential-

    $$E\{|h_l|^2\} = e^{\left(-\frac{l}{10}\right)}, \qquad l = 0, \ldots, L$$
  - Uniform-

    $$E\{|h_l|^2\} = 1, \qquad l = 0, \ldots, L$$
- Number of subcarriers per OFDM frame ($N$) = *64*
- Number of OFDM block used to estimate channel output auto-covariance matrix by time-averaging unless otherwise mentioned ($M$) = *1000*
- SNR (unless otherwise mentioned) = *30* dB
- The precoding matrix $\boldsymbol{W}$ is given by [10]:

$$\boldsymbol{W} = \boldsymbol{P}^{1/2}, \quad where\ \boldsymbol{P} = \begin{bmatrix} 1 & p & p & \cdots & p \\ p & 1 & p & \cdots & p \\ \vdots & \vdots & \vdots & \ddots & \vdots \\ p & p & p & \cdots & 1 \end{bmatrix}_{N \times N} \tag{3}$$

  The results are shown for precoding constant ($p$) = *0.5*
- The results are averaged over 25*0* Monte-Carlo runs.

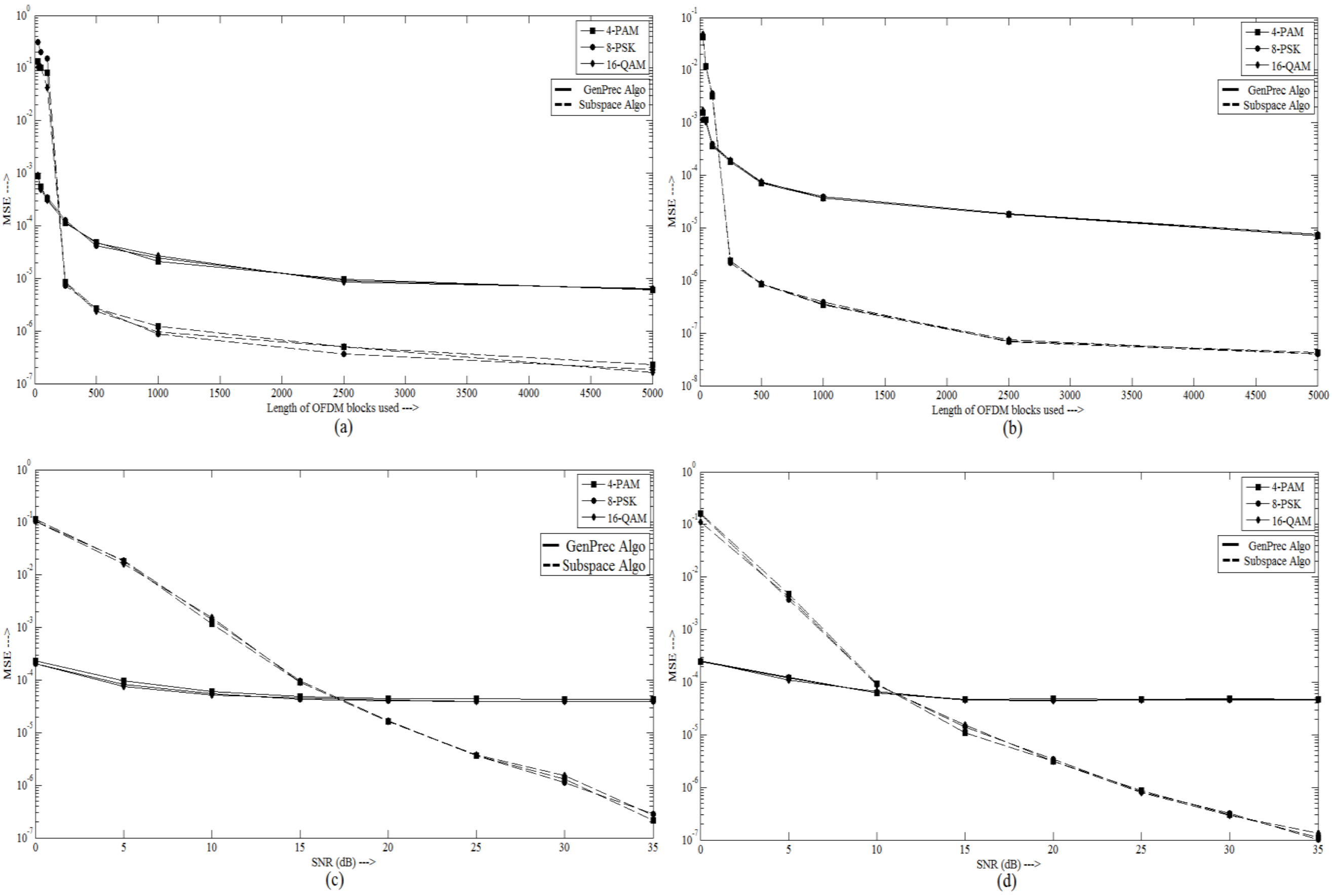

Figure 2: (a) MSE versus Length of OFDM blocks at *30* dB SNR for exponential channel PDP; (b) MSE versus Length of OFDM blocks at *30* dB SNR for uniform channel PDP; (c) MSE versus SNR (dB) with length of OFDM blocks equals to *1000* for exponential channel PDP and (d) MSE versus SNR (dB) with length of OFDM blocks equals to *1000* for exponential channel PDP

From Fig. 2, it can be observed that the performance is independent of the modulation scheme for both the techniques irrespective of channel PDP. The performance of both the techniques improves as the length of the OFDM blocks used is increased. The step-like characteristics of MSE performance is due to p. o. e. assumption (restriction) [7]. As stated by p. o. e. assumption, significant difference in performance of subspace algorithm can be observed around the threshold point. Unlike subspace-based method, the precoding-based technique shows gradual improvement in performance and provides a good estimate of the channel even when the length of OFDM frames is small. It can be seen that MSE value of order of $10^{-3}$ is achieved even when the length of OFDM blocks is as low as *25*. The performance of precoding-based technique is worse than that for subspace-based method when the length of OFDM blocks used exceeds the minimum threshold put forth by p. o. e. assumption. When the length of OFDM blocks is *1000*, the difference in the MSE performances of the two approaches is around *14* dB for exponential channel PDP case and *20* dB for uniform channel PDP case.

Over the range the SNR values from *0* dB to *35* dB, the performance improvement for the case of precoding-based method is *7* dB irrespective of channel PDP under consideration. In contrast to this, the subspace-based method shows an improvement of around *60* dB. For lower SNR values, the precoding-based technique performs better than the subspace-based method. The saturation observed in the performance of precoding-based technique is due to the fact that the channel estimates in this case are more sensitive to distortion due to non-availability of infinite OFDM blocks for time-averaging than distortion due to additive noise. In the case of precoding-based technique, at lower SNR (say *0* dB), the MSE performance is of the order of $10^{-3}$, which is at least *100* times better than that observed for subspace method. This is because, since the length of OFDM blocks used is *1000* and joint-correlation-averaging is used for channel estimation, the effect of AWGN is nullified. The 'cross-over' in the performance occurs at around *17* dB for exponential channel PDP case and at around *12* dB for uniform channel PDP case.

From the above results, it can be concluded that precoding-based technique provides a good estimate even when the length of OFDM frames available are small or when the SNR is low. This fact makes the precoding-technique a suitable candidate to low-power and high-mobility applications. The phase-ambiguity which is resolved using pilots is not feasible if the channel is fast time-varying. This calls for the need of blind techniques to resolve ambiguities.

## IV. COMPLETELY BLIND ALGORITHM

We propose a generalization of the constellation-splitting concept of [15] to make it less modulation dependent and develop a blind phase-estimation-correction algorithm applicable to any modulation schemes.

### A. *Generalized Constellation-Splitting*

The constellation-splitting can be done in any of the three ways:

a) The constellation is split into two halves, along x-axis (horizontal/ in-phase axis).
b) The constellation is split into two halves, along y-axis (vertical/ quadrature-phase axis).
c) The constellation is split into four halves, along both x-axis and y-axis.

We consider 16-QAM and illustrate the third case. This can be used without much alteration for PAM (one-dimensional version of QAM) and PSK (degenerative case of QAM).The concept of constellation-splitting and signal point assignment to subcarriers of OFDM frame for 16-QAM mapping is shown in Fig. 3 and Table 1. Each group of four subcarriers has any of the three unique phase points associated with them. It is reasonable to assume that the blind algorithm at the receiver has the knowledge of these phase values.

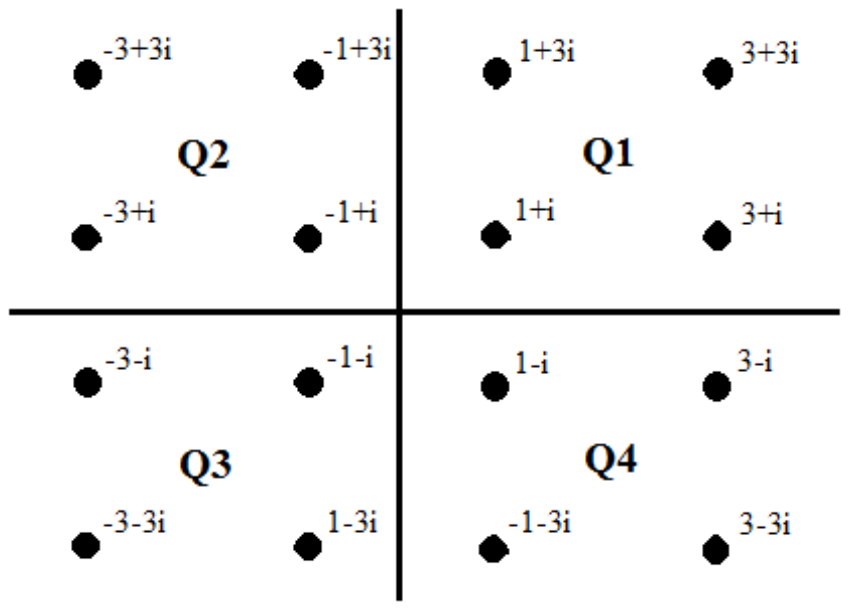

Figure 3: Signal constellation diagram of 16-QAM

TABLE 1: Phase-assignment scheme for Type c) constellation splitting technique

| **Quadrant number** | **Assigned to subcarriers** $k \rightarrow 0 - N\text{-}1$ **if** |
|---|---|
| 1 | $k$ mod $4 = 0$ |
| 2 | $(k\text{-}1)$ mod $4 = 0$ |
| 3 | $(k\text{-}2)$ mod $4 = 0$ |
| 4 | $(k\text{-}3)$ mod $4 = 0$ |

### B. *Phase-Correct Channel Estimation Algorithm*

Assume that the $n^{th}$ OFDM frame to be transmitted $\boldsymbol{d}(n)$ is constructed according to the rules of generalized constellation-splitting technique described above. The QAM signal points even though are from a finite alphabet set for a given quadrant, can be random within that quadrant (For example, four possible values of signal points exists per quadrant for 16-QAM). Hence, designing the precoder with phase-retaining property as proposed in [15] is not feasible. Thus, the phase estimates obtained via precoding-based technique cannot be used as an input for phase-ambiguity-estimator block. An alternative method to find the first approximation of the channel is described.

Assuming that the channel response remains more or less static for $T_0$ time units (spanning over $M$ OFDM frames), during the first $T_1$ ($T_1 < T_0$) units of time (spanning over $M_1$ OFDM frames) the transmitted OFDM frames are not precoded. The first approximation of the channel phase in time-domain is obtained at the receiver during this time by subspace method in time-domain using the redundancy induced by CP at the transmitter [16]. By taking N-point DFT of this estimate, the frequency-domain phase estimate of the channel which can be used as an input for phase-correction algorithm is obtained. The time duration $T_1$ is chosen such that subspace decomposition is feasible (see p. o. e. assumption in [7]). The channel amplitude estimates obtained using the subspace method is useless since they suffer from an ambiguity. During last $T_0$-$T_1$ units, the OFDM frames are precoded using the precoding matrix given by (3) with $p = 0.5$. The generalized precoding-based algorithm [10] is used to obtain channel's magnitude estimate in frequency-domain by tracing over different starting points ($q$). The amplitude and phase estimates thus obtained are combined to obtain the channel estimate which is of the form $\widehat{\mathbf{H}} = \mathbf{H}e^{j\varphi}$. Thus, a hybrid time-frequency algorithm is used to obtain the first estimate of the channel.

The phase-ambiguity estimate cannot be obtained directly by using the algorithm given in [15] since the source-phase-values and hence $\boldsymbol{B}$ is random from a given known set, unlike for the PAM case. Finding the bias vector involves a search over different phase values from the set. It can be concluded that, since the ambiguity term is theoretically constant over all the subcarriers (zero variance), for a given test vector, the ambiguity-estimate vector: $\widehat{\boldsymbol{\varphi}} = [\hat{\varphi}_0 \quad \cdots \quad \hat{\varphi}_{N-1}]^T$ has the least variance over the subcarriers if the phase values of test vector are same as that of the original source vector. This criterion is used as the search parameter. Further, since the ambiguity term is a constant over all the subcarriers; a subset of OFDM frame can be used for testing purpose. This reduces the number of test vector combinations that must be traced. For the present discussion, it is assumed that the test vector combinations are traced over first *4* subcarriers.

For 16-QAM, I-quadrant phase-value set is: $\boldsymbol{\varphi}_I = [18.4349^0 \ \ 45^0 \ \ 71.5651^0]$. Since the constellation is symmetric, the phase-value set for II, III and IV quadrant can be thus written as: $\boldsymbol{\varphi}_{II} = \boldsymbol{\varphi}_I + \pi/2$; $\boldsymbol{\varphi}_{III} = \boldsymbol{\varphi}_I + \pi$ and $\boldsymbol{\varphi}_{IV} = \boldsymbol{\varphi}_I + 3\pi/2$ respectively. For noisy-channel, the value of $\hat{\varphi}_k$ should be averaged over sufficient number of OFDM frames to combat the ill-effects of AWGN. For present discussion, the average is done over $M_1$ OFDM frames. Assuming that all the angles are in radians, the pseudo-code to obtain the ambiguity estimate is given as follows:

```
AmbiguityEstimate = [0 0 ... 0]^T
for Iter = 1 → M1
  Check = 1;
  for S1 = 1 → length(φI)
    for S2 = 1 → length(φII)
      for S3 = 1 → length(φIII)
        for S4 = 1 → length(φIV)
          BiasVector = [φI(S1) φII(S2) φII(S3) φII(S4)]^T
% Estimate of ambiguity-vector to be analysed for least-variance
          for k = 1 → 4
            ToAnalyse(k) = Ĥ(k) - y(k) + BiasVector(k)
          end
% Choose that source-phase vector as BiasVector which yields the lowest-variance-Ambiguity-Estimate vector
          if (Check = 1)
            AmbiguityEstimate (Iter) = mean (ToAnalyse)
            DecisionParameter = variance (ToAnalyse)
            Check = 0
          else
            if ((variance(ToAnalyse)) < DecisionParameter)
              AmbiguityEstimate (Iter) = mean (ToAnalyse)
              DecisionParameter = variance (ToAnalyse)
            end
          end
        end
      end
    end
  end
end
AmbiguityEstimate = mean (AmbiguityEstimate)   % Avg. over M1 frames
```

The proposed procedure is known as modified phase-directed (MPD) algorithm. This algorithm is similar to PD, but owing to constellation splitting which reduces the number of phase-values over which the search is to be performed and since phase-ambiguity is constant over all the subcarriers which facilitates the search over partial frame, MPD is computationally efficient. For a symmetric signal constellation, the error in the decision might occur when two or more source-phase vectors theoretically (noiseless case) yields zero-variance-ambiguity-vector estimate, or analogously in noisy channel case, the decision on the correct source-phase vector cannot be made with enough confidence. This condition occurs if and only if $Angle(\hat{\boldsymbol{s}}_1(n)) = Angle(\hat{\boldsymbol{s}}_2(n)) + K$, where $\hat{\boldsymbol{s}}_1(n)$ and $\hat{\boldsymbol{s}}_2(n)$ are two possible estimates of the source-phase vectors that yield zero-variance-ambiguity-vector estimate and $K$ is a constant angular-shift. The probability of occurrence of this situation is very low in practical scenarios. Further, this problem can be overcome by taking the average over number of OFDM frames.

## V. SIMULATION RESULTS FOR COMPLETELY BLIND ESTIMATOR

The simulation results to show the effectiveness of the proposed algorithm is shown in Fig. 4 and Fig. 5. The simulation parameters given in Section III are assumed except that the ambiguity is resolved using the proposed blind technique instead of pilots.

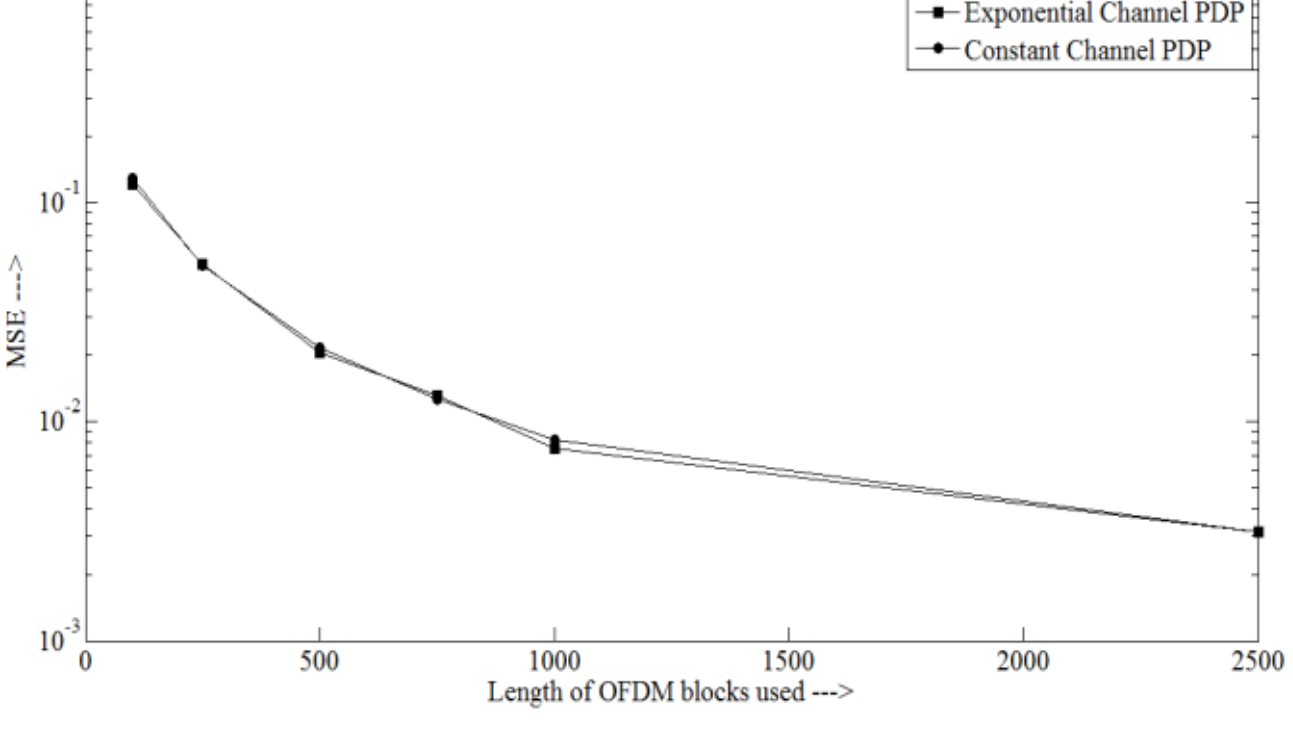

Figure 4: MSE versus Length of OFDM blocks at *25* dB SNR

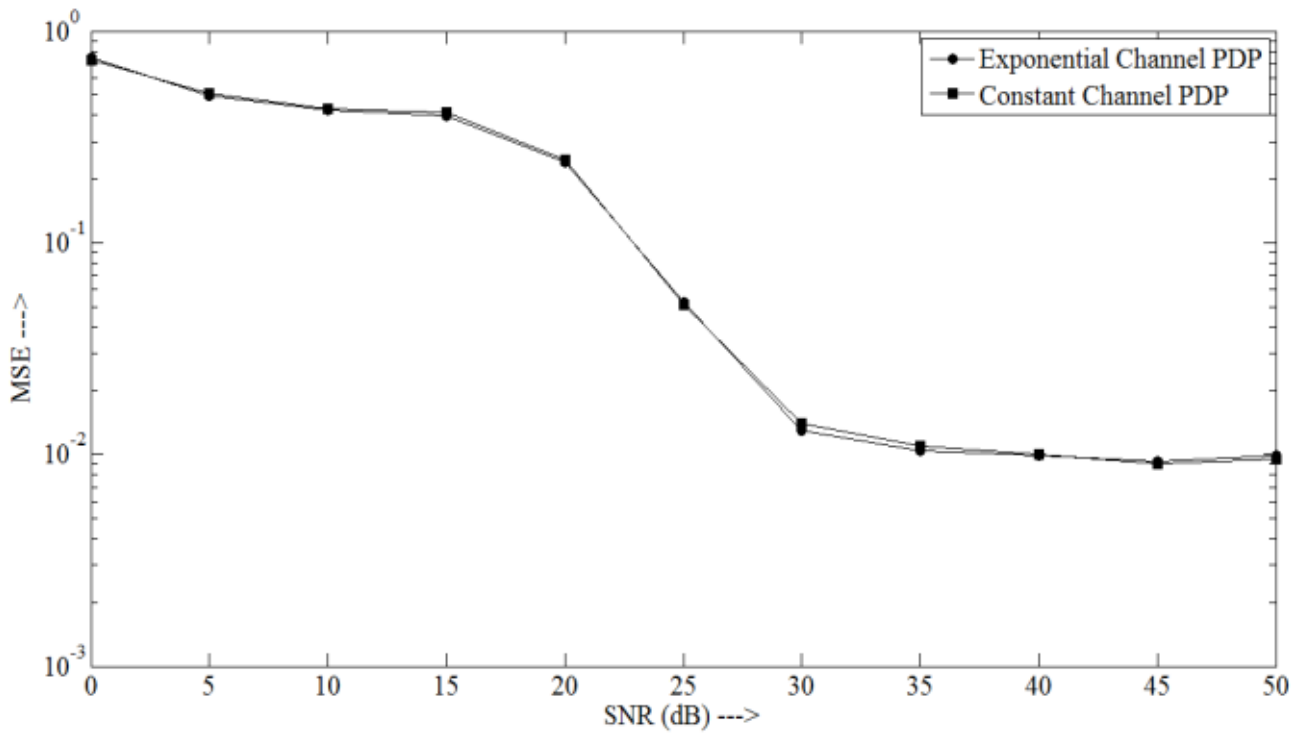

Figure 5: MSE versus SNR (dB) when length of OFDM blocks is 25*0*

It can be observed that the performance is independent of the channel PDP used. Since hybrid algorithm is used to obtain the first approximation of the channel, owing to p. o. e. assumption, the practical applicability of the estimator is restricted to relatively slow-varying channels. From Fig. 4, it can be observed that the MSE performance improves exponentially as the length of OFDM blocks is increased. The value of MSE is around $1.207\times10^{-1}$ when the length is *100* blocks and around $5.3\times10^{-2}$ when the length of OFDM blocks is *250*. When the length of OFDM blocks is *1000*, the MSE further reduces to around $8\times10^{-3}$. From Fig. 5, it can be observed that the MSE value is *0.73* at *0* dB and around *0.0572* at *25* dB. A linear improvement can be observed between *20* dB and *30* dB SNR values. The MSE settles at around $10^{-2}$ when the SNR is increased beyond 30 dB and after this point no further improvement is seen.

The performance of completely blind hybrid estimator is slightly worst than either of the two of its semi-blind counterparts: the precoding-based method and the subspace-based method at lower lengths of OFDM blocks or at lower SNR values but is comparable at higher ranges. It can be observed from Fig. 5 that at higher SNR values, the MSE performance of the estimator is acceptable for all practical purposes even when the length of OFDM blocks is *250*, the minimum threshold put forth by the p. o. e. assumption on subspace technique which is used to obtain first approximation of the channel phase. Further, the proposed algorithm is completely blind. The elimination of pilots/ reference symbols can be advantageous in fast time-varying channels.

## VI. CONCLUSION

The performance comparison of subspace- and precoding-based channel estimation techniques for SISO-OFDM systems is presented. The precoding-based technique is a suitable candidate for low power and high-mobility applications. These blind techniques suffer from an estimation ambiguity. Using precoding, the amplitude ambiguity can be resolved. In practice, the phase-ambiguity has been resolved by employing pilot carriers. We have proposed a completely blind channel estimation technique for SISO-OFDM using generalized constellation splitting and MPD algorithm. This algorithm has an advantage over its predecessors: the OFDM frame has zero-dc, the algorithm is structurally simpler and computationally efficient with an efficient encoding/ decoding logic. The performance analysis is presented for 16-QAM. It can be concluded from the simulation results that the completely blind techniques perform slightly worse than their semi-blind blind counterpart, the generalized precoding-based method [10] by around *15* dB, but yet, can estimate the channel within acceptable MSE tolerance for all practical purposes even when the length of blocks is low unlike subspace-based semi-blind method.